\documentclass[conference,a4paper,twocolumn]{IEEEtran}

\usepackage[english]{babel}
\usepackage{cite}

\usepackage{nicefrac,mathtools,units}
\usepackage{amsmath,amsfonts,amssymb,amsthm,bm}
\usepackage[caption=false,font=footnotesize]{subfig}
\usepackage{graphicx,booktabs}
\usepackage[boxed]{algorithm}
\usepackage{algorithmic,afterpage}
\usepackage{comment,todonotes}
\graphicspath{{./}{./fig/}}

\usepackage{tikz}
\usepackage{pgfplots}
\usepackage{tikz-3dplot}
\usetikzlibrary{arrows,shapes.geometric,calc,decorations.markings,patterns}
\definecolor{mathblue}{HTML}{3F3D9A}
\definecolor{mathpurp}{HTML}{9A3D71}
\definecolor{mathsand}{HTML}{9A8C3D}
\definecolor{mathgrn}{HTML}{0F9F4F}

\newtheorem{theorem}{Theorem}

\newtheorem{proposition}{Proposition}

\newtheorem{lemma}{Lemma}

\def\bs{\boldsymbol}
\def\bb{\mathbb}
\def\cl{\mathcal}

\def\ts{\textstyle}

\DeclareMathOperator{\diag}{{diag}}

\DeclareMathOperator{\Ex}{{\bb {E}}}
\DeclareMathOperator{\argmin}{{arg min}}

\def\eg{\emph{e.g.},~}

\def\ie{\emph{i.e.},~}

\def\ome{\boldsymbol{\omega}}
\def\eps{\boldsymbol{\varepsilon}}
\def\A{{\bm{A}}}
\def\a{\bm{a}}
\def\ail{\a_{i,l}}

\def\bxi{{\bm{\xi}}}
\def\B{\mathbb{B}}

\def\C{\bm{C}}
\def\Ciro{{\cl C}_\rho}

\def\d{\bm{d}}

\def\g{{\bm{\gamma}}}
\def\Bi{\bm{B}}
\def\gi{\bm{g}}

\def\H{{\cl H}}

\def\I{\bm{I}}

\def\Pro{\mathbb{P}}

\def\P{\bm{P}}
\def\onep{{{\bm 1}^\perp_m}}
\def\Ponep{\P_{\onep}}

\def\R{\mathbb{R}}

\def\s{\bm{s}}

\def\u{\bm{u}}

\def\vone{\bm{1}}
\def\vzer{\bm{0}}

\def\v{\bm{v}}

\def\x{{\bm{x}}}

\def\y{{\bm{y}}}

\newcommand{\tinv}[1]{\ts \tfrac{1}{#1}}

\pgfplotsset{
	compat=newest,
	ticklabel style = {font=\scriptsize},
	every axis label = {font=\small},
	legend style = {font=\small},
	label style = {font=\small},
	legend image code/.code={
		\draw[mark repeat=2,mark phase=2]
		plot coordinates {
			(0cm,0cm)
			(0.15cm,0cm)        
			(0.3cm,0cm)         
		};%
		}
}

\title{A Non-Convex Blind Calibration Method \\ for Randomised Sensing Strategies} 
\IEEEoverridecommandlockouts
\author{\IEEEauthorblockN{Valerio Cambareri and Laurent Jacques}
	\IEEEauthorblockA{Image and Signal Processing Group, ICTEAM/ELEN,\\
		Universit\'e catholique de Louvain,	Louvain-la-Neuve, Belgium.\\
		E-mail: {\{valerio.cambareri,~laurent.jacques\}@uclouvain.be}
		\thanks{The authors are funded by the Belgian F.R.S.-FNRS. Part of this study is funded by the project {\sc AlterSense} (MIS-FNRS).}
	}
}

\begin{document}
	\maketitle
	
	\begin{abstract}
		The implementation of computational sensing strategies often faces calibration problems typically solved by means of multiple, accurately chosen training signals, an approach that can be resource-consuming and cumbersome. Conversely, blind calibration does not require any training, but corresponds to a bilinear inverse problem whose algorithmic solution is an open issue. We here address blind calibration as a non-convex problem for linear random sensing models, in which we aim to recover an unknown signal from its projections on sub-Gaussian random vectors each subject to an unknown multiplicative factor (gain). To solve this optimisation problem we resort to projected gradient descent starting from a suitable initialisation. An analysis of this algorithm allows us to show that it converges to the global optimum provided a sample complexity requirement is met, \ie relating convergence to the amount of information collected during the sensing process. Finally, we present some numerical experiments in which our algorithm allows for a simple solution to blind calibration of sensor gains in computational sensing applications.
	\end{abstract}
	\begin{IEEEkeywords}
		Blind calibration, non-convex optimisation, sample complexity, computational sensing, bilinear inverse problems. 
	\end{IEEEkeywords}
	\section{Introduction}
	\label{sec:introduction}
	The problem of acquiring an unknown signal $\x$ in the presence of sensing model errors is crucial for modern computational sensing strategies such as Compressed Sensing (CS), in which such errors inevitably affect physical implementations and can significantly degrade signal recovery \cite{HermanStrohmer2010}. 
	Among the physical sources of such errors we may include: convolution kernels \cite{AhmedRechtRomberg2014,AhmedCosseDemanet2015,BahmaniRomberg2015a} as caused by lowpass optical elements, which affect the measured coefficients at the focal plane; attenuations and gains on the latter coefficients, \eg pixel response non-uniformity \cite{HayatTorresArmstrongEtAl1999}, fixed-pattern noise or vignetting; complex-valued gain and phase errors in sensor arrays \cite{FriedlanderStrohmer2014,LingStrohmer2015a,BilenPuyGribonvalEtAl2014}.
	
	Assuming such errors remain stationary throughout the sensing process, the use of linear random operators in CS suggests that repeating the acquisition, \ie taking several {\em snapshots} under new independent draws of a {\em randomised sensing model} could suffice to {diversify} the measurements and extract the information required to learn both the unknown signal and the model error. We here address the specific case of a single, unstructured vector $\x\in\R^n$ that is sensed by collecting $p$ snapshots of $m$ random projections, \ie our sensing model is
	\begin{equation}
	\label{eq:measurement-model-matrix}
	\y_{l} = \bar{\bs d}\,\A_{l} \x, \ \bar{\bs d} \coloneqq \diag(\d) \in \R^{m\times m}, \ l \in [p]\coloneqq\{1,\ldots,p\},
	\end{equation}  
	where $\y_l =  (y_{1,l}, \cdots, y_{m,l})^\top \in \R^m$ is the $l$-th snapshot; $\d = (d_1,\cdots,d_m)^\top \in \R^m_+$ is an unknown, positive and bounded gain vector that is identical throughout the $p$ snapshots; the random sensing matrices $\A_l\in \R^{m\times n}$ are independent and identically distributed (i.i.d.) and each $\A_l$ has i.i.d. rows, the $i$-th row $\a^\top_{i,l} \in \R^n$ being a centred isotropic (\ie $\Ex\a_{i,l} = \vzer_n, \Ex\ail \ail^\top=\I_n$) sub-Gaussian random vector (for a formal definition, see \cite[Section 5.2.5]{Vershynin2012}). 
	Note that $\x$ is also assumed to remain identical throughout $p$ snapshots, \eg a fixed scene being monitored by an imaging system.
	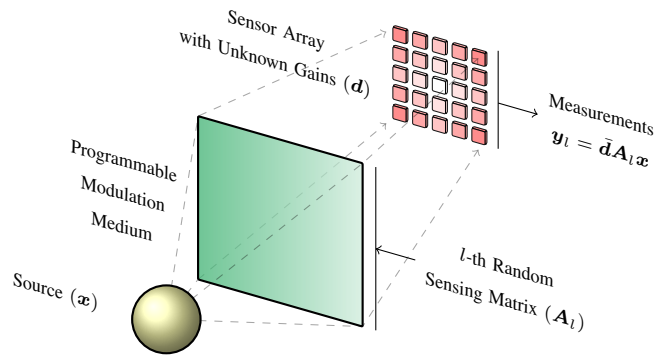
\begin{figure}	
		\centering
		\captionsetup{justification=centering}
		\tdplotsetmaincoords{60}{30}
		\begin{tikzpicture}[scale=1,tdplot_main_coords]
		\def\sx{0}
		\def\sy{1}
		\def\sz{0}
		\def\ps{0.3}
		\def\pps{0.2}
		\def\pys{0.05}
		\coordinate (SL) at (\sx,\sy,\sz);
		\coordinate (SS) at (0,-3,0);
		\coordinate (SX) at (0,-6,0);
		\begin{scope}
		\pgfmathsetseed{42}
		\foreach \tx in {-2,-1,...,2} {
			\foreach \ty in {-2,-1,...,2} {
				\pgfmathsetmacro{\tc}{25*(abs(\tx)/2+abs(\ty/2))}
				\draw[fill=red!\tc!white,rounded corners=0.5pt,opacity=0.9] ({\sx+\tx*\ps+\pps},{\sy},{\sz+\ty*\ps}) -- ({\sx+\tx*\ps+\pps},{\sy+\pys},{\sz+\ty*\ps}) -- ({\sx+\tx*\ps+\pps},{\sy+\pys},{\sz+\pps+\ty*\ps}) -- ({\sx+\tx*\ps+\pps},{\sy},{\pps+\sz+\ty*\ps}) -- cycle; 
				\draw[fill=red!\tc!white,rounded corners=0.5pt,opacity=0.9] ({\sx+\tx*\ps},\sy,{\pps+\sz+\ty*\ps}) -- ({\sx+\tx*\ps},{\sy+\pys},{\pps+\sz+\ty*\ps})
				-- ({\sx+\tx*\ps+\pps},{\sy+\pys},{\pps+\sz+\ty*\ps}) -- ({\sx+\tx*\ps+\pps},{\sy},{\pps+\sz+\ty*\ps}) -- cycle;
				\draw[fill=red!\tc!white,rounded corners=0.5pt,opacity=0.9] ({\sx+\tx*\ps},\sy,{\sz+\ty*\ps}) -- ({\sx+\tx*\ps+\pps},\sy,{\sz+\ty*\ps}) -- ({\sx+\tx*\ps+\pps},\sy,{\sz+\ty*\ps+\pps}) -- ({\sx+\tx*\ps},\sy,{\sz+\ty*\ps+\pps}) -- cycle;
			}
		}
		
		\coordinate (SC) at ({\sx+\pps/2},\sy,{\sz+\pps/2});
		\draw[->] ({\sx+\pps/2+3*\ps},\sy,{\sz+\pps/2}) -- +(0.6,0,0) node [sloped,near end,anchor=west, xshift=1ex,xslant=-0.25,align=center] {\scriptsize Measurements \\ \scriptsize $\y_l = \bar{\d} \A_l \x $};
		\draw[-] ({\sx+\pps/2+3*\ps},\sy,{\sz-2*\ps}) -- +(0,0,5*\ps);
		
		\def\tw{1.25}
		\draw[->,dashed,opacity=0.35] ($(SS)+(-\tw,0,-\tw)$) -> ({\sx-2*\ps-\pps/2},\sy,{\sz-2*\ps-\pps/2});
		
		\draw[black,thick,left color=mathgrn!80!white,right color=mathgrn!20!white,fill opacity=0.75,rounded corners=0.5pt] ($(SS)+(-\tw,0,-\tw)$) -- ($(SS)+(-\tw,0,\tw)$) -- ($(SS)+(\tw,0,\tw)$) -- ($(SS)+(\tw,0,-\tw)$) -- cycle; 
		\draw[-] ($(SS)+(\tw+\pps,0,-\tw)$) -- ($(SS)+(\tw+\pps,0,\tw)$);
		\draw[<-] ($(SS)+(\tw+\pps,0,0)$) -- +(0.6,0,0) node [sloped,near end,anchor=west, xshift=1ex,xslant=-0.25,align=center] {\scriptsize $l$-th Random \\ \scriptsize Sensing Matrix $(\A_l)$};
		\draw[draw=none] ($(SS)+(-2*\tw,0,\tw*2/5)$) -- ($(SS)+(-\tw,0,\tw*2/5)$) node[near start,sloped,below,xslant=-0.25,align=center,xshift=-1ex] {\scriptsize Programmable \\ \scriptsize Modulation \\ \scriptsize Medium};
		\draw[draw=none] ($(SL)+(-2*\tw,0,\tw/3)$) -- ($(SL)+(-\tw,0,\tw/3)$) node[near start,sloped,below,xslant=-0.25,align=center,xshift=-1ex,yshift=0.5ex] {\scriptsize Sensor Array \\ \scriptsize with Unknown Gains $(\d)$};	
		\draw[draw=none] ($(SX)+(-1.25*\tw,0,\tw/3)$) -- ($(SX)+(-\tw,0,\tw/3)$) node[near start,sloped,below,xslant=-0.25,align=center,yshift=-1ex,xshift=-1ex] {\scriptsize Source $(\x)$};
		
		
		\draw[->,dashed,opacity=0.35] (SX) -- ($(SS)+(\tw,0,\tw)$) -> ({\sx+2*\ps+\pps/2},\sy,{\sz+2*\ps+\pps/2});
		\draw[->,dashed,opacity=0.35] (SX) -- ($(SS)+(-\tw,0,\tw)$) -> ({\sx-2*\ps-\pps/2},\sy,{\sz+2*\ps+\pps/2});
		\draw[->,dashed,opacity=0.35] (SX) -- ($(SS)+(\tw,0,-\tw)$) -> ({\sx+2*\ps+\pps/2},\sy,{\sz-2*\ps-\pps/2});
		\draw[dashed,opacity=0.35] (SX) -- ($(SS)+(-\tw,0,-\tw)$);
		
		\shade[tdplot_screen_coords,ball color=yellow!30!white,draw=black,thick,opacity=1,fill opacity=1] (SX) circle(0.45) node {};
		
		\end{scope}
		\end{tikzpicture}
		\caption{\label{fig:modello} A randomised sensing model; {blind calibration} entails the joint recovery of the source $\x$ and sensor gains $\d$ by exploiting multiple random sensing matrices $\A_l$ (\eg $p$ programmable random masks in a random convolution setup \cite{Romberg2009}). The intensity of $\d$ is represented in shades of red as a possible {vignetting} of the sensor array.}
		
	\end{figure}
	
	\begin{table*}[!t]
		\centering
		\caption{\label{tab:quants} Finite-sample and expected values of the objective function; \newline its gradient components and Hessian matrix; the initialisation point $(\bxi_0,\g_0)$.}
		{
			\begin{tabular}{c c c}
				\toprule
				\bf Quantity &  \bf Finite-sample $(p < \infty)$ &  \bf Expectation $({\bb E}_{\ail}, p\rightarrow \infty)$ \\
				\toprule
				Objective Function: $f(\bxi,\g)$ & $\tfrac{1}{2mp} \sum_{l=1}^{p}
				\left\Vert \bar{\g} \A_l \bxi -
				\bar {\d} \A_l \x \right\Vert^2_2$ & ${\begin{gathered}\nonumber\tfrac{1}{2m}\left[\|\bxi\|^2_2 \|\g\|^2_2 + \|\x\|^2_2 \|\d\|^2_2 - 2 (\g^\top \d)(\bxi^\top \x)\right] \\ \equiv \tfrac{1}{2m} \|\bxi \g^\top - \x\d^\top\|^2_F\nonumber\end{gathered}}$ \\
				\midrule
				Signal Gradient: $\bs\nabla_{\bxi} f(\bxi,\g)$ & $\tfrac{1}{mp}
				\sum^p_{l=1} \A^\top_l \bar{\g} \left(\bar{\g} \A_l \bxi - \bar{\d} \A_l \x\right)$ & $\tfrac{1}{m}\left[\|\g\|^2_2 \bxi - (\g^\top \d) \x \right]$\\
				\addlinespace
				Gain Gradient: $\bs\nabla_{\g} f(\bxi,\g)$ & $\tfrac{1}{mp} \sum^p_{l=1}  \overline{\A_l \bxi} \left(\bar{\g} \A_l \bxi - \bar{\d} \A_l \x\right) $ & $\tfrac{1}{m} \left[\|\bxi\|^2_2 \g - (\bxi^\top\x) \d \right]$\\
				\addlinespace
				Projected Gain Gradient: $\bs\nabla^\perp_{\g} f(\bxi,\g)$ & $\tfrac{1}{mp} \sum^p_{l=1}  \Ponep \overline{\A_l \bxi} \left(\bar{\g} \A_l \bxi - \bar{\d} \A_l \x\right) $ & $\tfrac{1}{m} \left[\|\bxi\|^2_2 \eps - (\bxi^\top\x) \ome \right]$\\
				\midrule 
				Hessian Matrix: {$\H f(\bxi,\g)$}& {$\tfrac{1}{mp} \sum^p_{l=1} \bigg[\begin{smallmatrix} \A_l^\top \bar{\g}^2 \A_l & \A_l^\top\,\overline{2 \bar{\g}\A_l\bxi - \bar{\d}\A_l\x} \\ 
					\overline{2 \bar{\g}\A_l\bxi - \bar{\d}\A_l\x}\,\A_l & \overline{\A_l\bxi}^2
					\end{smallmatrix}\bigg]$} & {$\tfrac{1}{m} {\bigg[\begin{smallmatrix} \|\g\|^2_2 \I_n & 2 \bxi \g^\top - \x \d^\top \\ 2 \g \bxi^\top - \d \x^\top & \|\bxi\|^2_2 \I_m\end{smallmatrix}\bigg]}$}\\
				\midrule
				Initialisation: {$(\bxi_0,\,\g_0)$}& $\left(\tfrac{1}{mp} \sum^{p}_{l = 1} \left(\A_l\right)^\top \bar{\d}\,\A_l \x,\, \vone_m\right)$ & $\left(\tfrac{\|\d\|_1}{m}\x,\,\vone_m\right)$\\
				\bottomrule
			\end{tabular}
		}
		
	\end{table*}
	
	The inverse problem corresponding to this {\em bilinear} sensing model is hereafter called {\em blind calibration} \cite{BalzanoNowak2008,BilenPuyGribonvalEtAl2014}. In particular \eqref{eq:measurement-model-matrix} relates to computational sensing applications in which unknown $\d$ are associated to positive gains (see Figure \ref{fig:modello}) while $p$ random matrix instances can be applied on a source $\x$ by a suitable (\ie programmable) light modulation embodiment. This setup matches compressive imaging configurations \cite{Romberg2009,BjorklundMagli2013,DegrauxCambareriGeelenEtAl2014,BahmaniRomberg2015a,Dumas2016}~with an important difference in that the absence of {\em a priori} structure on $(\x,\d)$ in \eqref{eq:measurement-model-matrix} implies an over-Nyquist sampling regime with respect to (w.r.t.) $n$,~ \ie exceeding the number of unknowns as $m p \geq n + m$. When the effect of $\d$ is critical (\ie assuming $\bar{\d} \approx \I_m$ would lead to an inaccurate recovery of $\x$) solving this problem justifies a possibly over-Nyquist sampling regime as long as $(\x,\d)$ can both be recovered accurately (\eg as an on-line calibration). 
	
	Prior approaches to blind calibration entail solving convex or alternating optimisation algorithms  \cite{BalzanoNowak2008,BilenPuyGribonvalEtAl2014,LiporBalzano2014} aided by multiple input signals (\eg $\x_l, l \in [p]$) instead of taking new draws of the sensing operator itself. However, such methods lack formal recovery guarantees and require the training signals to be as independent as possible, or to lie in a low-dimensional subspace (possibly known {\em a priori}). More recently, {\em lifting} approaches \cite{LingStrohmer2015} have been proposed to jointly recover $(\x,\d)$ in \eqref{eq:measurement-model-matrix} (as well as for more general {\em blind deconvolution} problems \cite{AhmedRechtRomberg2014,LingStrohmer2015a,AhmedCosseDemanet2015,BahmaniRomberg2015a}). Their main limitation is in that a semidefinite program is solved to recover a large-scale rank-one matrix $\x \d^\top$; this approach becomes computationally inefficient and unaffordable quite rapidly as $m$ and $n$ exceed a few hundreds. 
	
	Inspired by recent results on fast, provably convergent non-convex algorithms for phase retrieval \cite{CandesLiSoltanolkotabi2015,WhiteSanghaviWard2015,SunQuWright2016} we address the recovery of $(\x,\d)$ by solving a non-convex problem presented in Section \ref{sec:ncbc}, as naturally defined by  \eqref{eq:measurement-model-matrix}. In particular, the use of $mp$ measurements in the model 
	allows us to find an unbiased estimator of $\x$ as $p \rightarrow \infty$; for $p < \infty$ we will initialise our algorithm with this estimator and run projected gradient descent 
	to obtain a recovery of $(\hat{\x},\hat{\d})$. 
	In Section \ref{sec:recguars} the properties of the gradient and initialisation under random sensing vectors $\ail$ will allow us to find a proof of convergence for this descent algorithm once a {\em sample complexity} requirement on the amount of collected measurements is met, \ie giving a bound on the number of snapshots that scales as $p \gtrsim \sqrt{m}$. 
	In Section \ref{sec:numexps} we provide numerical evidence on the algorithm's empirical phase transition; this is followed by a practical application of our method in a realistic computational sensing context.
	
	\section{A Non-Convex Approach to Blind Calibration}
	\label{sec:ncbc}
	\subsection{Problem Statement}
	The formulation of an inverse problem for \eqref{eq:measurement-model-matrix} is quite natural by means of a Euclidean data fidelity objective function $f(\bxi,\g) \coloneqq  \tfrac{1}{2mp} {\ts \sum_{l=1}^{p} \left\Vert \bar{\g} \A_l \bxi - \y_l\right\Vert^2_2}$, \ie we solve
	\begin{equation}%
	\label{eq:bcp}
	(\hat{\x}, \hat{\d}) = {\mathop{\argmin}_{\bxi \in \R^n, \g \in \Pi_+^m}} f(\bxi,\g),
	\end{equation}%
	given $\{\y_l\}^p_{l=1}, \{\A_l\}^p_{l=1}$, with $\Pi^m_+ \coloneqq \{\g \in \R^m_+, \,  \vone^\top_m \g = m\}$ being the scaled probability simplex. 
	To begin with, replacing \eqref{eq:measurement-model-matrix} in \eqref{eq:bcp}~shows that all points in 
	$
	\{(\bxi,\g) \in\R^n \times \R^m : \bxi= {\alpha}^{-1} \x, \g = \alpha \d, \alpha \in \R \setminus \{0\} \}
	$
	are global minimisers of $f(\bxi,\g)$ up to an unrecoverable scaling $\alpha$. In fact, the constraint $\g \in \Pi^m_+$ fixes one global minimiser $(\x^\star,\d^\star) = \big( \tfrac{\|\d\|_1}{m} \x, \tfrac{m}{\|\d\|_1}\d\big)$, \ie scaled by $\alpha = \nicefrac{m}{\|\d\|_1}$ (for insight on the identifiability of bilinear inverse problems we refer the reader to recent advancements, \eg \cite{Kech2016}). 
	In addition, by expanding the objective function $f(\bxi,\g)$ of \eqref{eq:bcp}, its gradient~$\bs\nabla f(\bxi,\g) = \left[\begin{smallmatrix}(\bs\nabla_\bxi f(\bxi,\g))^\top & (\bs\nabla_\g f(\bxi,\g))^\top\end{smallmatrix}\right]^\top$ and Hessian matrix $\H f(\bxi,\g)$ we can obtain the expressions reported in Table~\ref{tab:quants}. There, we confirm that $f(\bxi,\g)$ is generally non-convex (as noted in \cite{LingStrohmer2015} it is {\em biconvex}, \ie convex once either $\bxi$ or $\g$ are fixed) as there exist plenty of counterexamples $(\bxi',\g')$ for which the Hessian matrix $\H f(\bxi',\g') \nsucceq 0$. 	
	Table~\ref{tab:quants} also reports the case $p \rightarrow \infty$, where all finite-sample expressions are shown to be unbiased estimates of their expectation w.r.t.  the sensing vectors $\ail$. 
	
	To analyse problem \eqref{eq:bcp} further we proceed as follows. Letting $\d \in \R^m_+$ in \eqref{eq:measurement-model-matrix} be positive and bounded amounts to letting $\d^\star \in \Ciro \subset \Pi^m_+$, $\Ciro \coloneqq \vone_m + \onep \cap \rho\,\B^m_\infty$ for a maximum deviation $\rho \geq \|\d-\vone_m\|_\infty,\rho < 1$ which we assume known (with the orthogonal complement $\onep \coloneqq \{\v \in \R^m : \vone_m^\top\v= 0\}$ and $\B^q_p$ the $\ell_p$-ball in $\R^q$). 
	Thus, we can specify $\d^\star = \vone_m + \ome$ for $\ome \in \onep \cap	\rho\,\B^m_\infty, \rho < 1$, as well as $\g = \vone_m + \eps$ for $\eps \in \onep \cap	\rho\,\B^m_\infty$ provided that the algorithm solving \eqref{eq:bcp} will enforce $\g \in \Ciro$. This allows us to study \eqref{eq:bcp} in terms of the variations $\ome,\eps \in \onep \cap \rho \B^m_\infty$~around $\vone_m$ on the simplex $\Pi^m_+$. 
	
	While applying this constraint to the minimisation of $f(\bxi,\g)$ will not grant convexity in the domain of \eqref{eq:bcp}, we proceed by defining a {\em neighbourhood} of the global minimiser $(\x^\star,\d^\star)$ as follows. To begin with, we will require a notion of distance. To this end, we could adopt the pre-metric $${\Delta}_F(\bxi,\g) \coloneqq \tfrac{1}{m} \big\|\bxi \g^\top - \x \d^\top \big\|^2_F  \equiv 2 \Ex f(\bxi,\g),$$ the last equivalence being immediate from Table \ref{tab:quants}. While this is a naturally balanced definition, proofs with it are more cumbersome, so we resort to the simpler $${\Delta}(\bxi,\g) \coloneqq \|\bxi-\x^\star\|^2_2 + \tfrac{\|\x^\star\|^2_2}{m}\|\g-\d^\star\|^2_2,$$
	which for $(\bxi,\g) \in \R^n \times \Ciro, \rho \in (0,1)$ can be shown to verify $\ts(1\!-\!\rho){\Delta}(\bxi,\g)\leq{{\Delta}_F(\bxi,\g)} \leq(1\!+\!2\rho){\Delta}(\bxi,\g)$. With this we may define a neighbourhood of $(\x^\star,\d^\star)$ as
	$${\cl D}_{\kappa,\rho} \coloneqq \{(\bxi,\g)\in \R^n\times {\cl C}_\rho : \Delta(\bxi,\g) \leq \kappa^2 \|\x^\star\|^2_2\}, \ \rho \in [0, 1),$$
	that is the intersection of an ellipsoid in $\R^n \times \R^m$, as defined by $\Delta(\bxi,\g) \leq \kappa^2 \|\x^\star\|^2_2$ with $\R^n \times \Ciro$. 
	Rather than testing local convexity in a neighbourhood, we have found that a first-order analysis of $\bs\nabla f(\bxi,\g)$ on $(\bxi,\g) \in {\cl D}_{\kappa,\rho}$ suffices to prove our main results in Section \ref{sec:recguars}. 
	
	\subsection{Solution by Projected Gradient Descent}
	The solution of \eqref{eq:bcp} is here obtained as summarised in Algorithm \ref{alg1} and consists of an {\em initialisation} followed by projected gradient descent. Similarly to \cite{CandesLiSoltanolkotabi2015} we have chosen a signal-domain initialisation $\bxi_0$ that is an unbiased estimator of the exact solution as $p\rightarrow\infty$, \ie $\Ex\bxi_0 \equiv \x^\star$; this is indicated in Table~\ref{tab:quants}. For $p<\infty$ we will show in Proposition \ref{prop:init} that the initialisation lands in $(\bxi_0,\g_0) \in \cl D_{{\kappa},\rho}$ for $\rho \in [0, 1)$ with high probability, \ie there exists a sample complexity $mp$ ensuring ${\kappa}$ can be made sufficiently small. 
	
	As for the gains, we initialise $\g_0 = \vone_m \in \Pi^m_+ \ (\eps_0 = \vzer_m)$ and, since $\rho < 1$ is small, we perform a few simplifications to devise our solver to \eqref{eq:bcp}. While generally we would need to project any $\g$ on the simplex, we instead update the gains with the projected gradient $\bs\nabla^\perp_\g f(\bxi, \g) \coloneqq \Ponep \bs \nabla_\g f(\bxi, \g)$ (step 5:) using the projection matrix $\Ponep \coloneqq \I_m - \tfrac{1}{m} \vone_m \vone^\top_m$ (see Table \ref{tab:quants}). Then we apply $P_{\Ciro}$, \ie the projector on the convex set ${\Ciro}$ (step 6:). 		
	Actually, this step is just a formal requirement to ensure that each iterate $\g_{k+1} \in \Ciro \subset \Pi^m_+$ in proving the convergence of Algorithm \ref{alg1} to $(\x^\star,\d^\star)$; numerically, we have observed that step 6: can be omitted since $\underline{\g}_{k+1} \in \Ciro$ is always verified in our experiments. 
	
	Thus, Algorithm \ref{alg1} is a descent with the projected gradient $\bs\nabla^\perp f(\bxi,\g) \coloneqq \left[\begin{smallmatrix}(\bs\nabla_\bxi f(\bxi,\g))^\top & (\bs\nabla^\perp_\g f(\bxi,\g))^\top\end{smallmatrix}\right]^\top$; the proposed version performs two line searches in 3: that can be solved in closed-form at each iteration as 
	\begin{equation}
	\begin{gathered}
	\ts \mu_\bxi \coloneqq \frac{\sum^p_{l=1} \big\langle \bar{\g}_k \A_l \bxi_k,  \bar{\g}_k \A_l \bs\nabla_\bxi f(\bxi_k,\g_k) \big\rangle}{\sum^p_{l=1} \|\bar{\g}_k \A_l \bs\nabla_\bxi f(\bxi_k,\g_k)\|^2_2}\\
	\ts \mu_\g \coloneqq \frac{\sum^p_{l=1} \big\langle \overline{\A_l \bxi_k} \g_k,  \overline{\A_l \bxi_k} \bs\nabla^\perp_\g f(\bxi_k,\g_k) \big\rangle}{\sum^p_{l=1} \|\overline{\A_l \bxi_k} \bs\nabla^\perp_\g f(\bxi_k,\g_k)\|^2_2}
	\end{gathered}\label{eq:lsearches}
	\end{equation}
	and are simply introduced to improve the convergence rate. In the following we obtain the conditions that ensure convergence of this descent algorithm to the exact solution $(\x^\star,\d^\star)$ for some fixed step sizes $\mu_\bxi,\mu_\g$.
	
	\begin{algorithm}[!bt]
		{   \small
			\begin{algorithmic}[1]
				\STATE Initialise $\bxi_0 \coloneqq \tfrac{1}{m p}
				\sum^{p}_{l = 1} \left(\A_l\right)^\top \y_l,
				\, \g_0 \coloneqq \vone_m, \, k \coloneqq0$. 
				\WHILE{stop criteria not met}
				\STATE $\begin{cases} 
				\mu_\bxi \coloneqq \argmin_{\upsilon \in \R} f(\bxi_{k}-\upsilon\bs\nabla_\bxi f({\bxi_{k}, \g_{k}}), \g_{k}) \\
				\mu_\g \coloneqq \argmin_{\upsilon \in \R} f(\bxi_{k},\g_{k}-\upsilon\bs\nabla^\perp_\g f({\bxi_{k}, \g_{k}})) 
				\end{cases}$
				\STATE $\bxi_{k+1} \coloneqq \bxi_{k} - \mu_\bxi \bs\nabla_\bxi f({\bxi_{k}, \g_{k}})$
				\STATE $\underline{\g}_{k+1} \coloneqq \g_{k} - \mu_\g \, \bs\nabla^\perp_\g f({\bxi_{k}, \g_{k}})$ 
				\STATE ${\g}_{k+1} \coloneqq P_{\Ciro} \underline{\g}_{k+1}$
				\STATE $k \coloneqq k + 1$
				\ENDWHILE
			\end{algorithmic}
		}
		\caption{\label{alg1} Non-Convex Blind Calibration by Projected Gradient Descent.
			}
	\end{algorithm}
	
	\section{Convergence and Recovery Guarantees}
	\label{sec:recguars}
	Recalling that all $mp$ sensing vectors ${\ail}$ in \eqref{eq:measurement-model-matrix} are i.i.d. sub-Gaussian, we now establish the convergence of Algorithm \ref{alg1} in three steps: $(i)$ the initialisation $(\bxi_0,\g_0)$ is shown to lie in $\cl D_{{\kappa},\rho}$ for $\rho \in [0,1)$ and small $\kappa$ with high probability; $(ii)$ $\bs\nabla^\perp f(\bxi,\g)$ enjoys a condition by which, {\em uniformly} on $\cl D_{{\kappa},\rho}$, a gradient descent update decreases the distance to $(\x^\star,\d^\star)$; $(iii)$ by uniformity, applying this property to any $k$-th iterate $(\bxi_k,\g_k)$ leads to finding fixed step values $\mu_\bxi,\mu_\g$ that grant convergence to $(\x^\star,\d^\star)$ as $k\rightarrow \infty$. Proof sketches are provided after the main statements; the full arguments will be reported in an upcoming journal paper \cite{CambareriJacques2016}.
	
	We first state a key result and its application to proving the properties of our initialisation for $p<\infty$. 
	As typically done when deriving sample complexity bounds, we will use some universal constants $C,c>0$ changing from line to line.
	\begin{lemma}[Weighted Covariance Concentration Inequality]
		\label{lemma-1}
		Let $\{\ail \in \bb R^n: i \in [m], l\in[p]\}$ be a set of random vectors, each formed by $n$ i.i.d. copies of a sub-Gaussian random variable $X$ \cite[Section 5.2.3]{Vershynin2012} with $\Ex X = 0$, $\Ex X^2 = 1$ and sub-Gaussian norm $\|X\|_{\psi_2}$. For $\delta \in (0,1), t \geq 1$, provided $n \gtrsim t \log mp $ and $\ts mp \gtrsim \delta^{-2}(n+m)\log \tfrac{n}{\delta}$ we have, with probability exceeding 
		\begin{equation}
		\label{eq:pcoro}
		1 - C e^{- c \delta^2 m p} - (mp)^{-t}
		\end{equation} 
		for some $C,c>0$ depending only on $\|X\|_{\psi_2}$, that
		\begin{equation}
		\big\|\tinv{mp}\,\sum_{i=1}^m\ts \sum_{l=1}^p \theta_i (\ail \ail^\top - \I_n)\big\|_2 \leq
		\delta\,\|\bs \theta\|_\infty   
		\label{eq:devsnap}
		\end{equation}
		for all $\bs\theta = \{\theta_i\}^m_{i=1} \in \R^m$.
	\end{lemma}
	\begin{IEEEproof}[Proof sketch]
		By defining the function $$S(\u,\bs\theta)\coloneqq\tinv{mp}\sum_{i=1}^m\ts\sum_{l=1}^p \theta_i \big[(\u^\top \ail)^2-\|\u\|^2_2\big],$$ the proof consists in bounding $\sup_{\u\in{\bb S}^{n-1}_2} S(\u,\bs\theta)$ by a covering argument on $(\u,\bs\theta) \in {\bb S}^{n-1}_2 \times {\bb S}^{m-1}_\infty$ ($\bb S^{q-1}_p$ is the $\ell_p$-sphere in $\R^q$), using the concentration and continuity of $S(\u,\bs\theta)$.
	\end{IEEEproof}
	\begin{proposition}[Initialisation Proximity]
		\label{prop:init}
		Let $(\bxi_0,\g_0)$ be as in Table~\ref{tab:quants}. For any ${\delta} \in (0,1), t > 1$, provided $n \gtrsim t \log mp $ and $\ts mp \gtrsim {\delta}^{-2} (n + m) \log \tfrac{n}{\delta}$ we have, with probability exceeding 
		\[
		1 - C e^{- c {\delta}^2 m p} - (mp)^{-t}
		\]
		for some $C, c > 0$, that $\|\bxi_0 - \x^\star\|_2 \leq {\delta} \|\x^\star\|_2$. Since $\g_0 = \bs 1_m$ we also have $\|\g_0 - \d^\star\|_\infty\leq\rho<1$. Thus $(\bxi_0,\g_0) \in
		\cl D_{{\kappa},\rho}$ with the same probability and $\kappa \coloneqq \sqrt{\delta^2+\rho^2}$.
	\end{proposition}
	\begin{IEEEproof}[Proof sketch]
		Since $\bxi_0 = \tfrac{1}{mp} \sum^{p}_{l = 1} \left(\A_l\right)^\top \bar{\d}\,\A_l \x \equiv \tfrac{1}{mp}\ts\sum^m_{i=1} \sum^p_{l=1} d_i \ail \ail^\top \x$ we have
		\begin{align*}
		\|\bxi_0 - \x\|_2 \leq {\big\lVert\tfrac{1}{mp}\ts\sum^m_{i=1} \sum^p_{l=1} d_i (\ail \ail^\top - \I_n)\big\rVert_2} \lVert\x\rVert_2.
		\end{align*} 
		Using Lemma \ref{lemma-1} with some $\delta' \in (0,1)$ on the matrix norm at the right-hand side, and assigning ${\delta} \coloneqq \delta' (1+\rho)$ proves this Proposition.
	\end{IEEEproof}
	
	Secondly, we develop the requirements for convergence. Once the initialisation lies in $\cl D_{{\kappa},\rho}$, any update from $(\bxi,\g) \in \cl D_{{\kappa},\rho}$ to some $\bxi_+ \coloneqq \bxi - \mu_\bxi \bs\nabla_\bxi f({\bxi, \g}), \underline{\g}_+ \coloneqq \g - \mu_\g \bs\nabla^\perp_\g f({\bxi,\g})$ has distance from the solution
	\begin{align}
	& {\Delta} (\bxi_+,\underline{\g}_+) = {\Delta} (\bxi,\g)  \nonumber \\
	& \quad -2 \mu \big(\langle \bs\nabla_\bxi f(\bxi,\g), \bxi-\x^\star\rangle + \langle \bs\nabla^\perp_\g f(\bxi,\g), \g-\d^\star\rangle\big) \nonumber  \\ 
	& \quad + {\mu^2} \left(\|\bs\nabla_\bxi f(\bxi,\g)\|^2_2 + \tfrac{m}{\|\x^\star\|^2_2} \|\bs\nabla^\perp_\g f(\bxi,\g)\|_2^2\right) 
	\label{eq:updatedist}
	\end{align}		
	where we have let $\mu_\bxi \coloneqq \mu, \mu_\g \coloneqq \mu \tfrac{m}{\|\x^\star\|^2_2}$ for some $\mu > 0$. 
	To bound \eqref{eq:updatedist} we now verify a {\em regularity condition} on $\bs\nabla^\perp f(\bxi,\g)$ (analogue to
	\cite[Condition 7.9]{CandesLiSoltanolkotabi2015}) as a property holding uniformly on the neighbourhood with high probability.
	\begin{proposition}[Regularity condition in $\cl D_{{\kappa},\rho}$]
		\label{prop:regu}
		For any $\delta \in (0,1), t\geq1$, provided $\rho < \tfrac{1-2\delta}{9}$, $n \gtrsim t \log mp$, $p \gtrsim \delta^{-2} \log m$ and $\sqrt{m} p \gtrsim \delta^{-2} (n + m) \log \tfrac{n}{\delta}$ we have, with probability exceeding 
		\[
		1 - C \big[m e^{- c \delta^2 p} + e^{- c \delta^2 \sqrt{m} p} + e^{- c \delta^2 {m} p}  + (mp)^{-t}\big]
		\]
		for some $C,c>0$, that for all $(\bxi,\g) \in \cl D_{{\kappa},\rho}$
		\begin{align*}
		\label{eq:bounded-curvature}
		\left\langle \bs\nabla^\perp f(\bxi, \g), \left[\begin{smallmatrix} \bxi-\x^\star \\ \g-\d^\star \end{smallmatrix}\right] \right\rangle  \geq \tfrac{\eta}{2} \, {\Delta}(\bxi,\g), \tag{Bounded curvature}
		\end{align*}
		\begin{align*}
		\label{eq:bounded-lipschitz}        
		\|\bs\nabla^\perp f(\bxi, \g) \|^2_2 \tag{Lipschitz gradient} \leq
		L^2\ {\Delta}(\bxi,\g), 
		\end{align*}
		where $\eta \coloneqq 2(1-9\rho-2\delta)$, $L\coloneqq 4\sqrt{2}[1+\rho+(1+\kappa)\|\x^\star\|_2]$.
	\end{proposition}	
	\begin{IEEEproof}[Proof sketch]
		The general argument template uses triangle and Cauchy-Schwarz inequalities to manipulate the left-hand sides of the bounded curvature and \ref{eq:bounded-lipschitz} conditions (all required components are developed in Table~\ref{tab:quants}). Then, Lemma \ref{lemma-1}, a special version of it for $\bs\theta \in \vone^\perp_m$ that sets the requirement $\sqrt{m} p \gtrsim \delta^{-2} (n+m) \log \tfrac{n}{\delta}$, and \cite[Lemma 5.17]{Vershynin2012} allow us to bound terms of the type $\tinv{mp}\,\sum_{i=1}^m\ts \sum_{l=1}^p \theta_i \u^\top \ail \ail^\top \u$ for $\u \in \R^n$. 
		In more detail, reminding that $\g-\d^\star \equiv \eps-\ome$, the bounded curvature part of Proposition \ref{prop:regu} is shown to be
		\begin{align*}
		&\langle \bs \nabla_{\bxi} f(\bxi, \g), \bxi - \x^\star
		\rangle + \langle \bs \nabla^\perp_{\g} f(\bxi, \g), \g-\d^\star \rangle\\
		&\geq (1 - 5\rho)(1-\delta) \|\bxi - \x^\star\|^2_2+ \tinv{m}
		(1-\delta) \|\x^\star\|^2\|\eps - \ome\|_2^2 \\
		& \quad - 2(\delta + 4 \rho) \|\x^\star\|_2 \|\bxi - \x^\star\|_2
		\tfrac{\|\eps - \ome\|_2}{\sqrt m}\\
		&\geq (1 - 9\rho-{2}\delta) \Delta(\bxi,\g).
		\end{align*}
		Choosing $\eta \coloneqq 2 (1 - 9\rho-{2}\delta) > 0$ yields the condition on $\rho$. As for the \ref{eq:bounded-lipschitz} part,
		\begin{align*}
		&\|\bs \nabla_{\bxi} f(\bxi, \g)\|_2+ \|\bs \nabla^\perp_{\g} f(\bxi, \g)\|_2 \\
		&=\sup_{\u\in{\bb S}^{n-1}_2} \sup_{\v \in {\bb S}^{m-1}_2} \langle \bs \nabla_{\bxi} f(\bxi, \g), \u \rangle + \langle \bs \nabla^\perp_{\g} f(\bxi, \g), \v \rangle  \\
		&\leq (1+ \delta)(1+\rho)[1+\rho + (1+\kappa) \|\x^\star\|_2]\|\bxi-\x^\star\|_2 \\
		& \quad +  (1+ \delta)[1+\rho + (1+\kappa) \|\x^\star\|_2] \|\x^\star\|_2 \tfrac{\|\eps-\ome\|_2}{\sqrt m}.
		\end{align*}
		Thus, using straightforward inequalities on $\cl D_{\kappa,\rho}, {\rho} < 1$,
		\begin{align*}
		&\|\bs \nabla_{\bxi} f(\bxi, \g)\|^2_2 + \|\bs\nabla^\perp_{\g} f(\bxi, \g)\|^2_2 \nonumber \\
		&\leq 32 (1\!+\!\rho+(1\!+\!\kappa)\|\x^\star\|_2)^2 \Delta(\bxi,\g),
		\end{align*}	
		where we can collect $L \coloneqq 4\sqrt{2} (1+\rho+(1+\kappa)\|\x^\star\|_2)$ .	
	\end{IEEEproof}
	The obtained bounds allow for a proof of our main result. 
	\begin{theorem}[Provable Convergence to the Exact Solution]
		\label{th:1}
		Under the conditions of Proposition \ref{prop:init},
		\ref{prop:regu} we have that, with probability exceeding
		\[
		1 - C \big[m e^{- c \delta^2 p} + e^{- c \delta^2 \sqrt{m} p} + e^{- c {\delta}^2 m p}+ (mp)^{-t}\big],
		\]
		for some $C,c>0$, Algorithm \ref{alg1} with $\mu_\bxi \coloneqq \mu, \mu_\g \coloneqq \mu \tfrac{m}{\|\x^\star\|^2_2}$ has error decay
		\begin{equation}
		{\Delta}(\bxi_k,\g_k)\leq
		\big(1-\eta \mu + \tfrac{L^2}{\tau} \mu^2 \big)^k \big(\delta^2+\rho^2)\|\x^\star\|^2_2, {(\bxi_k, \g_k)\in {\cl D}_{\kappa,\rho}}
		\label{eq:convres}
		\end{equation}
		at any iteration $k > 0$  provided $\mu \in \big(0, \nicefrac{\tau \eta}{L^2} \big)$, $\tau \coloneqq \min\{1,\nicefrac{\|\x^\star\|^2_2}{{m}}\}$. Hence, ${\Delta}(\bxi_k,\g_k)\underset{k\rightarrow\infty}{\longrightarrow}0.$
	\end{theorem}
	\begin{IEEEproof}[Proof sketch]
		Assume Propositions \ref{prop:init}, \ref{prop:regu} jointly hold (hence the probability bound in this statement). For $k=0$, Proposition \ref{prop:init} grants ${\Delta}(\bxi_0,\g_0) \leq {\kappa}^2 \|\x^\star\|^2_2 + \tfrac{\|\x^\star\|^2_2}{m} m\rho^2$. Then, by Proposition \ref{prop:regu} and \eqref{eq:updatedist} we can bound
		\begin{align*}
		{\Delta}(\bxi_{k+1},\underline{\g}_{k+1}) & \equiv {\Delta}(\bxi_{k},\g_{k}) + \tfrac{\mu^2}{\tau} \|\bs\nabla^\perp f(\bxi_{k},\g_{k})\|^2_2 \nonumber \\ & \quad -2 \mu \left\langle \bs\nabla^\perp f(\bxi_{k},\g_{k}), \left[\begin{smallmatrix} \bxi_{k}-\x^\star \\ \g_{k}-\d^\star \end{smallmatrix}\right]\right\rangle \nonumber \\
		&\leq \big(1- \mu\eta + \tfrac{L^2}{\tau} \mu^2\big) {\Delta}(\bxi_{k},\g_{k})
		\end{align*}
		which decreases if we let $\mu \in \big(0, \nicefrac{\tau\eta}{L^2}\big)$. Moreover, since $P_{\Ciro}$ is {\em contractive} (since $\Ciro$ is a non-empty closed convex set) we have $\|\g_{k+1}-\d^\star\|_2\leq \|\underline{\g}_{k+1}-\d^\star\|_2$ and ${\Delta}(\bxi_{k+1},{\g}_{k+1}) \leq {\Delta}(\bxi_{k+1},\underline{\g}_{k+1})$. Applying recursively the inequalities for all $k > 0$ in Algorithm \ref{alg1} by uniformity on $\cl D_{{\kappa},\rho}$ yields \eqref{eq:convres}.
	\end{IEEEproof}
	We remark that the initialisation is critical to set the value of $\kappa$ in Propositions \ref{prop:init} and \ref{prop:regu}, with its value appearing in \eqref{eq:convres}. However, while the sample complexity for the initialisation is $mp \gtrsim (n+m)\log n$, ensuring convergence requires Proposition \ref{prop:regu} that sets $\sqrt{m} p \gtrsim (n+m)\log n$. Hence, our theory suggests the number of snapshots required for convergence is $p \gtrsim \sqrt{m} \log n$, even if our numerical experiments in Section \ref{sec:numexps} seem to indicate better rates than this bound.
	
	Finally, let us mention some extensions of our theory in which: $(i)$ a stability result can be obtained for Algorithm \ref{alg1} when \eqref{eq:measurement-model-matrix} is affected by bounded additive noise, which degrades gracefully the recovery quality of $\hat{\x},\hat{\d}$; $(ii)$ the signal $\x = \C \s$ and gains $\d = \Bi \gi$ depend on some low-dimensional parameters $\s,\gi$ with ${\rm dim}(\s) = k \ll n$, ${\rm dim}(\gi) = h \ll m$, thus improving the sample complexity which will scale as $k+h$. 
	These results will be presented in \cite{CambareriJacques2016}.
	
	\begin{figure}
		\centering
		\includegraphics{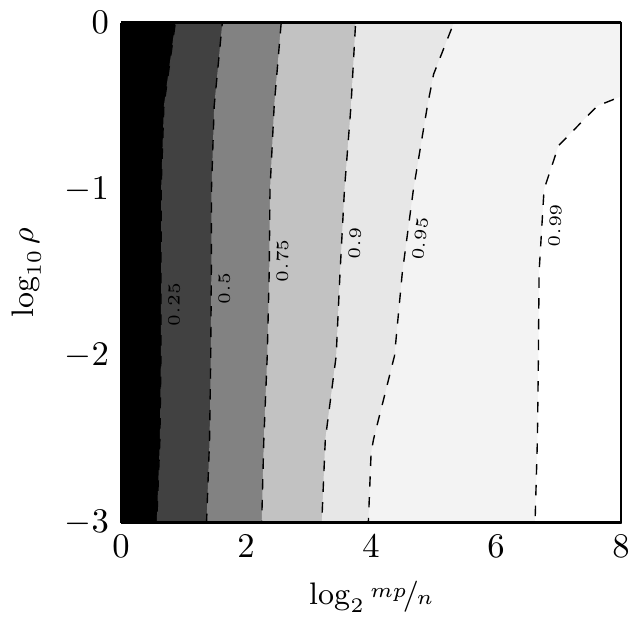}%
		
		\caption{\label{fig:pht} Empirical phase transition of \eqref{eq:bcp} solved by Algorithm \ref{alg1}. We report the contours $\{0.25,\ldots,0.99\}$ of the probability of exact recovery ${\rm P}_\zeta({mp},\rho)$.}
		
	\end{figure}
	\section{Numerical Experiments}
	\label{sec:numexps}
	\subsection{Empirical Phase Transition}
	To trace the empirical phase transition of Algorithm \ref{alg1} we ran some extensive simulations by
	generating $144$ random instances of \eqref{eq:measurement-model-matrix}, fixing $n = 2^8, m=2^6$ and varying $p = \{2^2, \ldots, 2^{10}\},\rho = \{10^{-3}, \ldots, 1\}$. Each instance was drawn with $\x \in \B^n_2$, $\ail\sim \cl N(\vzer_n,\I_n)$, $\d = \vone_m + \ome $ with $\ome \in \onep \cap \rho \,\bb S^{m-1}_\infty$. Then, we solved \eqref{eq:bcp} by our descent algorithm and evaluated ${\rm P_\zeta} \coloneqq \Pro\left[\ts\max\left\{\tfrac{\|\hat{\d}-\d^\star\|_2}{\|\d^\star\|_2},\tfrac{\|\hat{\x}-\x^\star\|_2}{\|\x^\star\|_2}\right\}
	<\zeta\right]$ on the trials with $\zeta = \unit[-70]{dB}$  (in accordance with the stop criterion at $f(\bxi_k,\g_k) < 10^{-7}$). The results are reported in Figure \ref{fig:pht}, in which we highlight the contour levels of $\rm P_\zeta$.
	
	\begin{figure}
		\vspace{-8pt}
		\null\hfill
		\subfloat[{$\x$ (true signal)}]{
			\includegraphics[width=1in]{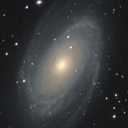}
		}
		\hfill
		\subfloat[{$\hat{\x}$ recovered by LS with model error}]{
			\includegraphics[width=1in]{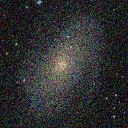}
		}
		\hfill
		\subfloat[{$\hat{\x}$ recovered by \eqref{eq:bcp}}]{
			\includegraphics[width=1in]{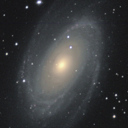}
		}
		\hfill\null			
		\\[-1ex]
		\subfloat[{$\d$ (true sensor gains), $\rho \approx 0.99$}]{				
			\includegraphics{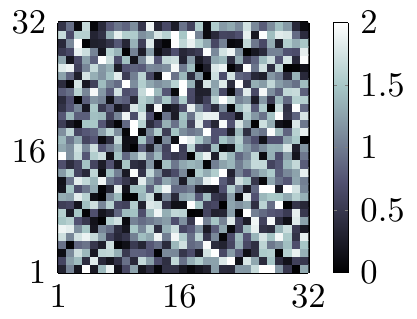}
		}
		\subfloat[{$\alpha \hat{\d}$ recovered by \eqref{eq:bcp}}]{
			\includegraphics{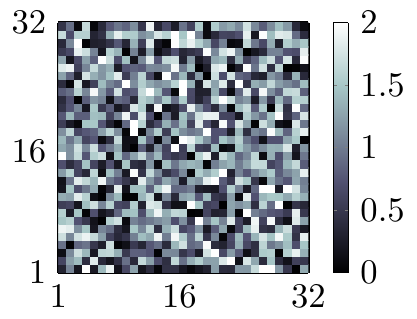}
		}
		\hfill\null
		\vspace{-1ex}
		\caption{\label{fig:randexp} A high-dimensional example of blind calibration for computational sensing; the unknown gains $\d$ $(m=\unit[32\times32]{pixel})$ and signal $\x$ $(n = \unit[128\times128]{pixel})$ are perfectly recovered with $p = 32$ snapshots.}
		
	\end{figure}%
	\subsection{Blind Calibration of a Randomised Imaging System}
	To test  our approach in a realistic context we assume that $\x$ is a $n=\unit[128\times128]{pixel}$ colour image acquired by a sensing device that implements \eqref{eq:measurement-model-matrix} in which its $m=\unit[32\times32]{pixel}$ sensor array suffers from a randomly drawn attenuation profile $\d \in \Pi^m_+$ generated as before, fixing $\rho\approx0.99$. We capture $p=32$ (\ie $mp = 2n$) snapshots with $\ail\sim \cl N(\vzer_n,\I_n)$ (each colour channel is processed separately). By running Algorithm \ref{alg1} we obtain the results depicted in Figure \ref{fig:randexp}; the recovered $(\hat{\x},\hat{\d})\equiv({\x},{\d})$ by solving \eqref{eq:bcp} attains $\ts\max\left\{\tfrac{\|\hat{\d}-\d^\star\|_2}{\|\d^\star\|_2},\tfrac{\|\hat{\x}-\x^\star\|_2}{\|\x^\star\|_2}\right\} \approx \unit[-61.59]{dB}$ in accordance with the stop criterion at $f(\bxi_k,\g_k) < 10^{-6}$. 
	Instead, by fixing $\g \coloneqq \vone_m$ and solving \eqref{eq:bcp} only w.r.t. $\bxi$, \ie finding the least-squares (LS) solution $\hat{\x}$ and fully suffering the model error, we obtain $\tfrac{\|\hat{\x}-\x^\star\|_2}{\|\x^\star\|_2} \approx \unit[-5.99]{dB}$. 	
	Moreover, Figure \ref{fig:decay} reports the evolution of the distances $\Delta(\bxi_k,\g_k)$ and $\Delta_F(\bxi_k,\g_k)$ measured on two runs of this exemplary case with the same stop criterion, one with the line searches \eqref{eq:lsearches} (ending at $k=369$) and one with a fixed step $\mu_\bxi = \mu \coloneqq 10^{-4},\mu_\g \coloneqq \mu \tfrac{m}{\|\bxi_0\|^2_2} \approx 0.88$ (ending at $k=6301$). We observe that there is clearly a linear bound on the convergence rate and that \eqref{eq:lsearches} clearly achieves faster convergence than a fixed-step choice of $\mu_\bxi,\mu_\g$. 
	Finally we note that while the theory in Section \ref{sec:recguars} is developed for sub-Gaussian $\A_l$ in \eqref{eq:measurement-model-matrix}, this experiment can be shown to run successfully when $\A_l$ is implemented (\eg optically) as a random convolution \cite{Romberg2009}. 
	
	\section{Conclusion}
	\label{sec:conclusion}
	We presented and solved a non-convex formulation of blind calibration for linear random sensing models affected by unknown gains. In absence of {\em a priori} structure on the signal and gains, the sample complexity required for convergence must verify $\sqrt{m} p \gtrsim (n+m)\log n$. Future developments of this approach include the extension to complex gains (\ie $\d \in \bb C^m$), as well as modifying the algorithm to enforce the sparsity of $\x$ (or $\d$) by which a reduction of the sample complexity below $n + m$ (\ie for actual CS) will be obtained.
	\afterpage{
		\begin{figure}[!t]
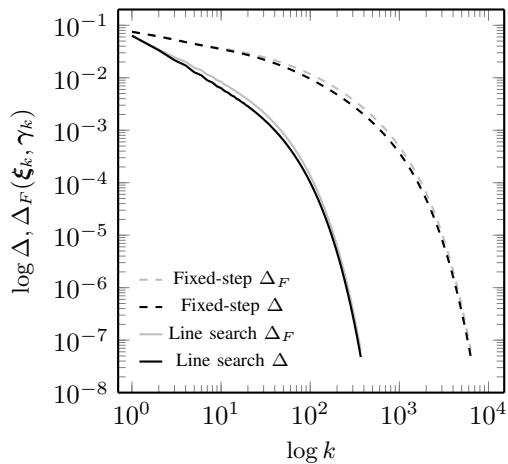

			\centering
%
			
			\caption{\label{fig:decay} Residual evolution of two exemplary runs of Algorithm \ref{alg1}: with fixed steps $\mu_\bxi,\mu_\g$ (dashed); with the line search updates given in \eqref{eq:lsearches} (solid).}
			
		\end{figure}
	}
	\bibliographystyle{IEEEtran}
	\bibliography{cosera_blind_2016}

\begin{thebibliography}{10}
\providecommand{\url}[1]{#1}
\csname url@samestyle\endcsname
\providecommand{\newblock}{\relax}
\providecommand{\bibinfo}[2]{#2}
\providecommand{\BIBentrySTDinterwordspacing}{\spaceskip=0pt\relax}
\providecommand{\BIBentryALTinterwordstretchfactor}{4}
\providecommand{\BIBentryALTinterwordspacing}{\spaceskip=\fontdimen2\font plus
\BIBentryALTinterwordstretchfactor\fontdimen3\font minus
  \fontdimen4\font\relax}
\providecommand{\BIBforeignlanguage}[2]{{%
\expandafter\ifx\csname l@#1\endcsname\relax
\typeout{** WARNING: IEEEtran.bst: No hyphenation pattern has been}%
\typeout{** loaded for the language `#1'. Using the pattern for}%
\typeout{** the default language instead.}%
\else
\language=\csname l@#1\endcsname
\fi
#2}}
\providecommand{\BIBdecl}{\relax}
\BIBdecl

\bibitem{HermanStrohmer2010}
M.~A. Herman and T.~Strohmer, ``{General deviants: An analysis of perturbations
  in compressed sensing},'' \emph{IEEE Journal of Selected Topics in Signal
  Processing}, vol.~4, no.~2, pp. 342--349, 2010.

\bibitem{AhmedRechtRomberg2014}
A.~Ahmed, B.~Recht, and J.~Romberg, ``Blind deconvolution using convex
  programming,'' \emph{IEEE Transactions on Information Theory}, vol.~60,
  no.~3, pp. 1711--1732, 2014.

\bibitem{AhmedCosseDemanet2015}
A.~Ahmed, A.~Cosse, and L.~Demanet, ``A convex approach to blind deconvolution
  with diverse inputs,'' in \emph{2015 IEEE 6\textsuperscript{th} International
  Workshop on Computational Advances in Multi-Sensor Adaptive Processing
  (CAMSAP)}, Dec. 2015, pp. 5--8.

\bibitem{BahmaniRomberg2015a}
S.~Bahmani and J.~Romberg, ``{Lifting for Blind Deconvolution in Random Mask
  Imaging: Identifiability and Convex Relaxation},'' \emph{SIAM Journal on
  Imaging Sciences}, vol.~8, no.~4, pp. 2203--2238, 2015.

\bibitem{HayatTorresArmstrongEtAl1999}
M.~M. Hayat, S.~N. Torres, E.~Armstrong, S.~C. Cain, and B.~Yasuda,
  ``Statistical algorithm for nonuniformity correction in focal-plane arrays,''
  \emph{Applied Optics}, vol.~38, no.~5, pp. 772--780, 1999.

\bibitem{FriedlanderStrohmer2014}
B.~Friedlander and T.~Strohmer, ``Bilinear compressed sensing for array
  self-calibration,'' in \emph{2014 48\textsuperscript{th} {Asilomar}
  {Conference} on {Signals}, {Systems} and {Computers}}, Nov. 2014, pp.
  363--367.

\bibitem{LingStrohmer2015a}
S.~Ling and T.~Strohmer, ``Self-calibration and biconvex compressive sensing,''
  \emph{Inverse Problems}, vol.~31, no.~11, p. 115002, 2015.

\bibitem{BilenPuyGribonvalEtAl2014}
C.~Bilen, G.~Puy, R.~Gribonval, and L.~Daudet, ``Convex {Optimization
  Approaches} for {Blind Sensor Calibration Using Sparsity},'' \emph{IEEE
  Transactions on Signal Processing}, vol.~62, no.~18, pp. 4847--4856, Sep.
  2014.

\bibitem{Vershynin2012}
R.~Vershynin, ``Introduction to the non-asymptotic analysis of random
  matrices,'' in \emph{Compressed Sensing: Theory and Applications}.\hskip 1em
  plus 0.5em minus 0.4em\relax Cambridge University Press, 2012, pp. 210--268.

\bibitem{Romberg2009}
J.~Romberg, ``Compressive sensing by random convolution,'' \emph{SIAM Journal
  on Imaging Sciences}, vol.~2, no.~4, pp. 1098--1128, 2009.

\bibitem{BalzanoNowak2008}
L.~Balzano and R.~Nowak, ``{Blind calibration of networks of sensors: Theory
  and algorithms},'' in \emph{Networked Sensing Information and Control}.\hskip
  1em plus 0.5em minus 0.4em\relax Springer, 2008, pp. 9--37.

\bibitem{BjorklundMagli2013}
T.~Bjorklund and E.~Magli, ``A parallel compressive imaging architecture for
  one-shot acquisition,'' in \emph{2013 IEEE Picture Coding Symposium
  (PCS)}.\hskip 1em plus 0.5em minus 0.4em\relax IEEE, 2013, pp. 65--68.

\bibitem{DegrauxCambareriGeelenEtAl2014}
K.~Degraux, V.~Cambareri, B.~Geelen, L.~Jacques, G.~Lafruit, and G.~Setti,
  ``{Compressive Hyperspectral Imaging by Out-of-Focus Modulations and
  Fabry-P{\'e}rot Spectral Filters},'' in \emph{International Traveling
  Workshop on Interactions between Sparse models and Technology (iTWIST)},
  2014.

\bibitem{Dumas2016}
J.~P. Dumas, M.~A. Lodhi, W.~U. Bajwa, and M.~C. Pierce, ``Computational
  imaging with a highly parallel image-plane-coded architecture: challenges and
  solutions,'' \emph{Opt. Express}, vol.~24, no.~6, pp. 6145--6155, Mar 2016.

\bibitem{LiporBalzano2014}
J.~Lipor and L.~Balzano, ``Robust blind calibration via total least squares,''
  in \emph{Acoustics, Speech and Signal Processing (ICASSP), 2014 IEEE
  International Conference on}.\hskip 1em plus 0.5em minus 0.4em\relax IEEE,
  2014, pp. 4244--4248.

\bibitem{LingStrohmer2015}
S.~Ling and T.~Strohmer, ``{Blind Deconvolution Meets Blind Demixing:
  Algorithms and Performance Bounds},'' \emph{arXiv preprint arXiv:1512.07730},
  2015.

\bibitem{CandesLiSoltanolkotabi2015}
E.~Cand\`es, X.~Li, and M.~Soltanolkotabi, ``Phase {Retrieval} via {Wirtinger
  Flow}: {Theory} and {Algorithms},'' \emph{IEEE Transactions on Information
  Theory}, vol.~61, no.~4, pp. 1985--2007, Apr. 2015.

\bibitem{WhiteSanghaviWard2015}
C.~D. White, S.~Sanghavi, and R.~Ward, ``The local convexity of solving systems
  of quadratic equations,'' \emph{arXiv:1506.07868 [math, stat]}, Jun. 2015,
  arXiv: 1506.07868.

\bibitem{SunQuWright2016}
J.~Sun, Q.~Qu, and J.~Wright, ``A geometric analysis of phase retrieval,'' in
  \emph{2016 IEEE International Symposium on Information Theory (ISIT)}, July
  2016, pp. 2379--2383.

\bibitem{Kech2016}
M.~Kech and F.~Krahmer, ``Optimal injectivity conditions for bilinear inverse
  problems with applications to identifiability of deconvolution problems,''
  \emph{arXiv preprint arXiv:1603.07316}, 2016.

\bibitem{CambareriJacques2016}
V.~Cambareri and L.~Jacques, ``{Through the Haze: A Non-Convex Approach to
  Blind Calibration for Linear Random Sensing Models},'' 2016, in preparation.

\end{thebibliography}
\end{document}